\documentclass{aa}

\usepackage{amsmath}
\usepackage{txfonts}              % following A&A author'S Guide - 3 January 2008, p. 7
\usepackage{natbib}
\usepackage{lscape}               % following ftp://ftp.edpsciences.o<g/pub/aa/readme.html
\usepackage[usenames]{color}
\usepackage{wasysym}            % \jupiter
\usepackage{marvosym}           % @
\usepackage{footnote}
\usepackage{footmisc}

\definecolor{darkblue}{rgb}{0,0,.4}

\long\def\symbolfootnote[#1]#2{\begingroup\def\thefootnote{\fnsymbol{footnote}}\footnote[#1]{#2}\endgroup}

\bibpunct{(}{)}{;}{a}{}{,}             % to follow the A&A style

\title{Transit detections of extrasolar planets around main-sequence stars}

\subtitle{I. Sky maps for Hot Jupiters\thanks{Sky maps (Figures 1 and 3) can be downloaded in electronic form at the CDS via anonymous ftp to cdsarc.u-strasbg.fr (130.79.128.5) or via http://cdsweb.u-strasbg.fr/cgi-bin/qcat?J/A+A/.}}

\author{R. Heller \inst{1} \and D. Mislis \inst{2} \and J. Antoniadis \inst{3}}

\institute{
Hamburger Sternwarte (Universit\"at Hamburg), Gojenbergsweg 112, 21029 Hamburg, Germany\\
\email{rheller@hs.uni-hamburg.de}
\and
Hamburger Sternwarte (Universit\"at Hamburg), Gojenbergsweg 112, 21029 Hamburg, Germany\\
\email{dmislis@hs.uni-hamburg.de}
\and
Aristotle University of Thessaloniki, Dept. of Physics, Section of Astrophysics, Astronomy and Mechanics, GR-541\,24 Thessaloniki, Greece\\
\email{iantonia@physics.auth.gr}
}

\date{Received 24 April 2009 / Accepted 03 October 2009}

\abstract
{The findings of more than 350 extrasolar planets, most of them nontransiting Hot Jupiters, have revealed correlations between the metallicity of the main-sequence (MS) host stars and planetary incidence. This connection can be used to calculate the planet formation probability around other stars, not yet known to have planetary companions. Numerous wide-field surveys have recently been initiated, aiming at the transit detection of extrasolar planets in front of their host stars. Depending on instrumental properties and the planetary distribution probability, the promising transit locations on the celestial plane will differ among these surveys.}
{We want to locate the promising spots for transit surveys on the celestial plane and strive for absolute values of the expected number of transits in general. Our study will also clarify the impact of instrumental properties such as pixel size, field of view (FOV), and magnitude range on the detection probability.}
{We used data of the Tycho catalog for $\approx~1$ million objects to locate all the stars with $0^\mathrm{m}~\lesssim~m_\mathrm{V}~\lesssim~11.5^\mathrm{m}$ on the celestial plane. We took several empirical relations between the parameters listed in the Tycho catalog, such as distance to Earth, $m_\mathrm{V}$, and $(B-V)$, and those parameters needed to account for the probability of a star to host an observable, transiting exoplanet. The empirical relations between stellar metallicity and planet occurrence combined with geometrical considerations were used to yield transit probabilities for the MS stars in the Tycho catalog. Magnitude variations in the FOV were simulated to test whether this fluctuations would be detected by BEST, XO, SuperWASP and HATNet.}
{We present a sky map of the expected number of Hot Jupiter transit events on the basis of the Tycho catalog. Conditioned by the accumulation of stars towards the galactic plane, the zone of the highest number of transits follows the same trace, interrupted by spots of very low and high expectation values. The comparison between the considered transit surveys yields significantly differing maps of the expected transit detections. While BEST provides an unpromising map, those for XO, SuperWASP, and HATNet show FsOV with up to 10 and more expected detections. The sky-integrated magnitude distribution predicts 20 Hot Jupiter transits with orbital periods between 1.5\,d and 50\,d and $m_\mathrm{V}~<~8^\mathrm{m}$, of which two are currently known. In total, we expect 3412 Hot Jupiter transits to occur in front of MS stars within the given magnitude range. The most promising observing site on Earth is at latitude $~=~-1$.}
{}


\begin{document}

\titlerunning{Transit detections of extrasolar planets -- I.}

\authorrunning{R. Heller et al.}

\maketitle

\section{Introduction}
\label{sec:intro}

A short essay by Otto Struve \citep{1952Obs....72..199S} provided the first published proposal of transit events as a means of exoplanetary detection and exploration. Calculations for transit detection probabilities \citep{1971Icar...14...71R, 1984Icar...58..121B, 2006AcA....56..183P} and for the expected properties of the discovered planets have been done subsequently by many others \citep{2005A&A...442..731G, 2007A&A...475..729F, 2008ApJ...686.1302B}. Until the end of the 1990s, when the sample of known exoplanets had grown to more than two dozen \citep{2000ApJ...532L..51C}, the family of so-called `Hot Jupiters', with 51 Pegasi as their prototype, was unknown and previous considerations had been based on systems similar to the solar system. Using geometrical considerations, \citet{1971Icar...14...71R}\footnote{A correction to his Eq. (2) is given in \citet{1984Icar...58..121B}.} found that the main contribution to the transit probability of a solar system planet would come from the inner rocky planets. However, the transits of these relatively tiny objects remain undetectable around other stars as yet.

The first transit of an exoplanet was finally detected around the sun-like star HD209458 \citep{2000ApJ...529L..45C, 2000A&A...359L..13Q}. Thanks to the increasing number of exoplanet search programs, such as the ground-based Optical Gravitational Lensing Experiment (OGLE) \citep{1992AcA....42..253U}, the Hungarian Automated Telescope (HAT) \citep{2002PASP..114..974B, 2004PASP..116..266B}, the Super Wide Angle Search for Planets (SuperWASP) \citep{2003ASPC..294..405S}, the Berlin Exoplanet Search Telescope (BEST) \citep{2004PASP..116...38R}, XO \citep{2005PASP..117..783M}, the Transatlantic Exoplanet Survey (TrES) \citep{2007ASPC..366...13A}, and the Tautenburg Exoplanet Search Telescope (TEST) \citep{2009IAUS..253..340E} and the space-based missions `Convection, Rotation \& Planetary Transits' (CoRoT) \citep{2002ESASP.485...17B} and Kepler \citep{2007CoAst.150..350C}, the number of exoplanet transits has grown to 62 until September $1^\mathrm{st}$ 2009\footnote{Extrasolar Planets Encyclopedia (EPE): www.exoplanet.eu. Four of these 62 announced transiting planets have no published position.} and will grow drastically within the next years. These transiting planets have very short periods, typically $<~10$\,d, and very small semimajor axes of usually $<~0.1$\,AU, which is a selection effect based on geometry and Kepler's third law \citep{1619QB41.K38.......}. Transiting planets with longer periods present more of a challenge, since their occultations are less likely in terms of geometrical considerations and they occur less frequently.

Usually, authors of studies on the expected yield of transit surveys generate a fictive stellar distribution based on stellar population models. \citet{2007A&A...475..729F} use a Monte-Carlo procedure to synthesize a fictive stellar field for OGLE based on star counts from \citet{2006AcA....56....1G}, a stellar metallicity distribution from \citet{2004A&A...418..989N}, and a synthetic structure and evolution model of \citet{2003A&A...409..523R}. The metallicity correlation, however, turned out to underestimate the true stellar metallicity by about 0.1\.dex, as found by \citet{2004A&A...415.1153S} and \citet{2005ApJ...622.1102F}. In their latest study, \citet{2009A&A...504..605F} first generate a stellar population based on the Besan\c{c}on catalog from \citet{2003A&A...409..523R} and statistics for multiple systems from \citet{1991A&A...248..485D} to apply then the metallicity distribution from \citet{2004A&A...415.1153S} and issues of detectability \citep{2006MNRAS.373..231P}. \citet{2008ApJ...686.1302B} rely on a Galactic structure model by \citet{1980ApJS...44...73B}, a mass function as suggested by \citet{2002AJ....124.2721R} based on Hipparcos data, and a model for interstellar extinction to estimate the overall output of the current transit surveys TrES, XO, and Kepler. In their paper on the number of expected planetary transits to be detected by the upcoming Pan-STARRS survey \citep{2004SPIE.5489...11K}, \citet{2009A&A...494..707K} also used a Besan\c{c}on model as presented in \citet{2003A&A...409..523R} to derive a brightness distribution of stars in the target field and performed Monte-Carlo simulations to simulate the occurrence and detections of transits. These studies include detailed observational constraints such as observing schedule, weather conditions, and exposure time and issues of data reduction, e.g. red noise and the impact of the instrument's point spread function.

In our study, we rely on the extensive data reservoir of the Tycho catalog instead of assuming a stellar distribution or a Galactic model. We first estimate the number of expected exoplanet transit events as a projection on the complete celestial plane. We refer to recent results of transit surveys such as statistical, empirical relationships between stellar properties and planetary formation rates. We then use basic characteristics of current low-budget but high-efficiency transit programs (BEST, XO, SuperWASP, and HATNet), regardless of observational constraints mentioned above, and a simple model to test putative transits with the given instruments. With this procedure, we yield sky maps, which display the number of expected exoplanet transit detections for the given surveys, i.e. the transit sky as it is seen through the eyeglasses of the surveys.

The Tycho catalog comprises observations of roughly 1 million stars taken with the Hipparcos satellite between 1989 and 1993 \citep{1997yCat.1239....0E, 1997ESASP.402...25H}. During the survey, roughly 100 observations were taken per object. From the derived astrometric and photometric parameters, we use the right ascension ($\alpha$), declination ($\delta$), the color index $(B-V)$, the apparent visible magnitude $m_\mathrm{V}$, and the stellar distance $d$ that have been calculated from the measured parallax. The catalog is almost complete for the magnitude limit $m_\mathrm{V}~\lesssim~11.5^\mathrm{m}$, but we also find some fainter stars in the list.

\section{Data analysis}
\label{sec:data_anal}

The basis of our analysis is a segmentation of the celestial plane into a mosaic made up of multiple virtual fields of view (FsOV). In a first approach, we subdivide the celestial plane into a set of $181~\times~361~=~65\,341$ fields. Most of the current surveys do not use telescopes, which typically have small FsOV, but lenses with FsOV of typically $8^{\circ}~\times~8^{\circ}$. Thus, we apply this extension of $8^{\circ}~\times~8^{\circ}$ and a stepsize of $\Delta\delta~=~1^\circ~=~\Delta\alpha$, with an overlap of $7^\circ$ between adjacent fields, for our automatic scanning in order to cover the complete sky. We chose the smallest possible step size in order to yield the highest possible resolution and the finest screening, despite the high redundancy due to the large overlap. A smaller step size than $1^\circ$ was not convenient due to limitations of computational time. An Aitoff projection is used to fold the celestial sphere onto a 2D sheet.

\subsection{Derivation of the stellar parameters}
\label{sub:derivation}

One key parameter for all of the further steps is the effective temperature $T_\mathrm{eff}$ of the stars in our sample. This parameter is not given in the Tycho catalog but we may use the stellar color index $(B-V)$ to deduce $T_\mathrm{eff}$ by

\begin{equation}\label{equ:Teff_BV}
  T_\mathrm{eff} = 10^{[14.551 - (B-V)] / 3.684}\,\mathrm{K} \hspace{0.2cm} ,
\end{equation}

\noindent
which is valid for main-sequence (MS) stars with $T_\mathrm{eff} \lesssim~9\,100$\,K as late as type M8 \citep{1998JRASC..92...36R}. Although we apply this equation to each object in the catalog, of which a significant fraction might exceed $T_\mathrm{eff}~=~9\,100$\,K, this will not yield a serious challenge since we will dismiss these spurious candidates below. From the object's distance to Earth $d$ and the visible magnitude $m_\mathrm{V}$, we derive the absolute visible magnitude $M_\mathrm{V}$ via

\begin{equation}
  M_\mathrm{V} = m_\mathrm{V} - 5^{\mathrm{m}} \log \left(\frac{d}{10\,\mathrm{pc}}\right) \hspace{0.2cm} ,
\end{equation}

\noindent
where we neglected effects of stellar extinction. In the next step, we compute the stellar radius $R_\star$ in solar units via

\begin{equation}
  \frac{R_\star}{R_\odot} = \left[ \left( \frac{5\,770\,\mathrm{K}}{T_\mathrm{eff}} \right)^4 10^{(4.83 - M_\mathrm{V})/2.5} \right]^{1/2}
\end{equation}

\noindent
and the stellar mass $M_\star$ by

\begin{equation}\label{equ:mlr}
  M_\star = (4 \pi R_\star^2 \sigma_{\mathrm{SB}} T_\mathrm{eff}^4)^{1 / \beta} \hspace{0.2cm} ,
\end{equation}

\noindent
where $\sigma_{\mathrm{SB}}$ is the Stefan-Boltzmann constant. The coefficient $\beta$ in the relation $L~\propto~M^\beta$ depends on the stellar mass. We use the values and mass regimes that were empirically found by \citet{1983Ap&SS..96..125C}, which are listed in Table \ref{tab:beta} \citep[see also][]{1983Obs...103...29S}.

\renewcommand{\arraystretch}{1.3}
\begin{table}[h]
  \centering
  \caption{Empirical values for $\beta$ in the mass-luminosity relation Eq. (\ref{equ:mlr}) as given in \citet{1983Ap&SS..96..125C}.}
  \label{tab:beta}

    \begin{tabular}{l|c}

    \hline

    \hline

    \hline \hline

   {\sc $\beta$} & {\sc Stellar Mass Regime}\\
    \hline

    $3.05 \pm 0.14$ & $M_\star \lesssim 0.5\,M_\odot$  \\

    $4.76 \pm 0.01$ & $0.6\,M_\odot \lesssim M_\star \lesssim 1.5\,M_\odot$  \\

    $3.68 \pm 0.05$ & $1.5\,M_\odot \lesssim M_\star$  \\

    \hline
  \end{tabular}
\end{table}

We deduce the stellar metallicity $\mathrm{[Fe/H]}_\star$ from the star's effective temperature $T_\mathrm{eff}$ and its color index $(B-V)$ by

\begin{equation}\label{equ:Fe_H}
  \mathrm{[Fe/H]}_\star = \frac{1}{411} \left( \frac{T_\mathrm{eff}}{\mathrm{K}} - 8\,423 + 4\,736\,(B-V) - 1\,106\,(B-V)^2 \right) \hspace{0.18cm} ,
\end{equation}

\noindent
as given in \citet{2004A&A...415.1153S}. This relation, however, is only valid for stars with $0.51~<~(B-V)~<~1.33$, $4\,495\,\mathrm{K}~<~T_\mathrm{eff}~<~6\,339\,\mathrm{K}$, $-0.7~<~\mathrm{[Fe/H]}_\star~<~0.43$, and $\log(g)~>~4$. We reject those stars from the sample that do not comply with all these boundary conditions. On the one hand we cleanse our sample of non-MS stars, on the other hand the sample is reduced seriously. While our original reservoir, our `master sample', consists of 1\,031\,992 stars from the Tycho catalog, all the restrictions mentioned above diminish our sample to $ 392\,000$ objects, corresponding to roughly 38\,\%.

\section{Transit occurrence and transit detection}
\label{sec:trans_occ_detect}

\subsection{Transit occurrence}
\label{sub:occurance}

Now that we derived the fundamental stellar parameters, we may turn towards the statistical aspects of planetary occurrence, geometric transit probability and transit detection. We start with the probability for a certain star of the Tycho catalog, say the $i^\mathrm{th}$ star, to host an exoplanet. For F, G, and K dwarfs with $-0.5~<~\mathrm{[Fe/H]}_i~<~0.5$, \citet{2005ApJ...622.1102F} found the empirical relationship

\begin{equation}\label{equ:metallicity}
  \wp_{\mathrm{\exists \, planet},i} = 0.03 \cdot 10^{2 \cdot \mathrm{[Fe/H]}_i}
\end{equation}

\noindent
for a set of 850 stars with an analysis of Doppler measurements sufficient to detect exoplanets with radial velocity (RV) semiamplitudes $K~>~30\,\mathrm{ms}^{-1}$ and orbital periods shorter than 4\,yr (see \citet{2005PThPS.158...24M} and \citet{2007MNRAS.380.1737W} for a discussion of the origin of this formula and its implications for planet formation). These periods are no boundary conditions for our simulations since we are only interested in surveys with observing periods of $\leq~50$\,d. The additional constraint on the metallicity does not reduce our diminished sample of 392\,000 stars since there is no star with $-0.7~<~\mathrm{[Fe/H]}_i~<~-0.5$ in the Tycho catalog. Similar to the correlation we use, \citet{2007ARA&A..45..397U} found a metallicity distribution of exoplanet host stars equivalent to $\wp_{\mathrm{\exists \, planet},i}~=~0.044~\cdot~10^{2.04 \cdot \mathrm{[Fe/H]}_i}$. However, this fit was restricted to stars with $\mathrm{[Fe/H]}_\star~>~0$ since they suspect two regimes of planet formation. \citet{2009ApJ...697..544S} extended the uniform sample of \citet{2005ApJ...622.1102F} and found the power-law $\wp_{\mathrm{\exists \, planet},i} = 1.3 \cdot 10^{2 \cdot \mathrm{[Fe/H]}_i} + C$, $C \in \{0, 0.5\}$, to yield the best data fit. These recent studies also suggest that there exists a previously unrecognized tail in the planet-metallicity distribution for $\mathrm{[Fe/H]}_\star~<~0$. Taking Eq. (\ref{equ:metallicity}) we thus rather underestimate the true occurrence of exoplanets around the stars from the Tycho catalog. The metallicity bias of surveys using the RV method for the detection of exoplanets is supposed to cancel out the bias of transit surveys \citep{2006AcA....56....1G,2008ApJ...686.1302B}.

In the next step, we analyze the probability of the putative exoplanet to actually show a transit. Considering arbitrary inclinations of the orbital plane with respect to the observer's line of sight and including Kepler's third law, \citet{2000ApJ...545L..47G} found the geometric transit probability to be

\begin{equation}\label{equ:P_geo}
  \wp_{\mathrm{geo},i} = 23.8 \left( \frac{M_i}{M_\odot} \right)^{-1/3} \left( \frac{R_i}{R_\odot} \right) \left( \frac{P}{\mathrm{d}}\right)^{-2/3} \hspace{0.2cm} ,
\end{equation}

\noindent
where $P$ is the orbital period. A more elaborate expression -- including eccentricity, planetary radius, the argument of periastron and the semi-major axis instead of the orbital period -- is given by \citet{2003PASP..115.1355S}. Note that $\wp_{\mathrm{geo},i}$ in Eq. (\ref{equ:P_geo}) does not explicitly but implicitly depend on the semi-major axis $a$ via $P~=~P(a)$! The probability for an exoplanetary transit to occur around the $i^\mathrm{th}$ star is then given by

\begin{equation}\label{equ:P_occ}
  \wp_{\mathrm{occ},i} = \wp_{\mathrm{\exists \, planet},i} \cdot \wp_{\mathrm{geo},i} \hspace{0.2cm} ,
\end{equation}

\noindent
where $P$ is the remaining free parameter, all the other parameters are inferred from the Tycho data. Since we are heading for the expectation value, i.e. the number of expected transits in a certain field of view (FOV), we need a probability density for the distribution of the orbital periods of extrasolar planets. On the basis of the 233 exoplanets listed in the EPE on July $6^\mathrm{th}$ 2007, \citet{2007AJ....134.2061J} used a power-law fit $\delta(P)~=~C(k)~\cdot~(P/\mathrm{d})^{-k}$, with $C(k)$ as the normalization function, and the boundary condition for the probability density $\int_0^\infty~\mathrm{d}P~\delta(P)~=~1$ to get

\begin{equation}\label{equ:period_density}
  \delta(P) = \frac{1-k}{B^{1-k} - A^{1-k}} \left( \frac{P}{\mathrm{d}} \right)^{-k}
\end{equation}

\noindent
with $A~=~1.211909$\,d and $B~=~4517.4$\,d as the lower and upper limits for the period distribution and $k~=~0.9277$. This function is subject to severe selection effects and bases on data obtained from a variety of surveys and instruments. It overestimates short-period planets since \citet{2007AJ....134.2061J} included transiting planets and the associated selection effects. While the function presumably does not mirror the true distribution of orbital periods of exoplanets, it is correlated to the period distribution to which current instruments are sensitive, in addition to geometric selection effects as given by Eq. (\ref{equ:P_geo}).

We now segment the celestial plane into a mosaic made up of multiple virtual FsOV, as described at the beginning of Sect. \ref{sec:data_anal}, to calculate the number of expected transits in that field. In Sect. \ref{sub:detection} we will attribute the FOV of the respective instrument to that mosaic and we will also consider the CCD resolution. The number of stars comprised by a certain FOV is $n$. The number of expected transits around the $i^{\mathrm{th}}$ star in that field, $N_i$, with periods between $P_1$ and $P_2$ is then given by

\begin{align}\label{equ:Ni_expect}
  N_i = & \int_{P_1}^{P_2} \mathrm{d}P \ \delta(P) \ \wp_{\mathrm{occ},i} \\ \nonumber
      = & \ \wp_{\mathrm{\exists \, planet},i} \cdot 23.8 \left( \frac{M_i}{M_\odot} \right)^{-1/3} \left( \frac{R_i}{R_\odot} \right) \frac{1-k}{B^{1-k}-A^{1-k}} \\ \nonumber
          & \ \times \ \frac{1}{1/3-k} \left( P_2^{1/3-k} - P_1^{1/3-k} \right) \mathrm{d}^{2/3} \ \ _{| \ A < P_1 < P_2 < B}  \hspace{0.25cm},
\end{align}

\noindent
and the number of expected transits in the whole FOV is

\begin{equation}\label{equ:N_field_expect}
  N = \sum_{i=1}^n N_i \hspace{0.2cm} .
\end{equation}

\noindent
We emphasize that this is not yet the number of expected transit detections within a certain FOV (see Sect. \ref{sec:results}) but the number of expected transits to occur within it.

\begin{figure}
  \centering
  \scalebox{0.54}{\includegraphics{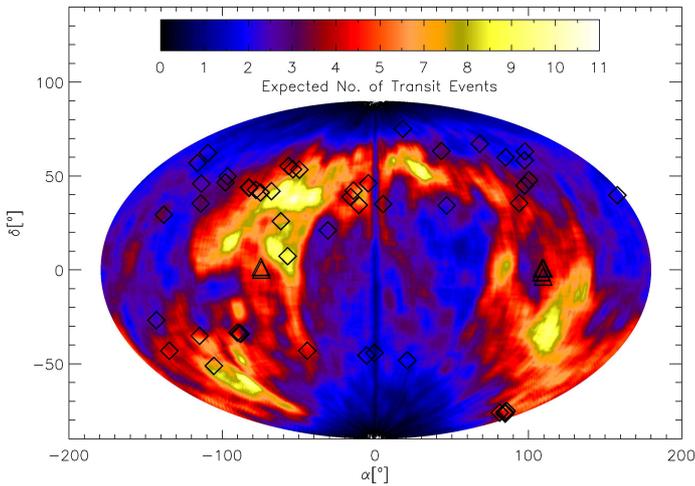}}
  \hspace{0.cm}
  \caption{Sky map of the expected number of exoplanet transit events, $N$, with orbital periods between $P_1~=~1.5$\,d and $P_2~=~50$\,d on the basis of 392\,000 objects from the Tycho catalog. The published positions of 58 transiting planets from the EPE as of September $1^\mathrm{st}$ 2009 are indicated with symbols: 6 detections from the space-based CoRoT mission are labeled with triangles, 52 ground-based detections marked with squares. The axes only refer to the celestial equator and meridian.}
  \label{fig:N_50d}
\end{figure}

A graphical interpretation of this analysis is presented in Fig. \ref{fig:N_50d}, where we show a sky map of the expected number of exoplanet transits around MS stars with $m_\mathrm{V}~\lesssim~11.5^\mathrm{m}$ for orbital periods between $P_1~=~1.5$\,d and $P_2~=~50$\,d. This map bases on several empirical relationships and on substantial observational bias towards close-in Jupiter-like planets, but nevertheless it represents the transit distribution to which current instrumentation has access to. The pronounced bright regions at the upper left and the lower right are the anti-center and the center of the Milky Way, respectively. The absolute values of $0.5~\lesssim~N~\lesssim~5$ for the most of the sky are very well in line with the experiences from wide-field surveys using a $6^\circ~\times~6^\circ$ field. \citet{2005ApJ...621.1061M} stated 5 to 20 or more exoplanet transit candidates, depending on Galactic latitude, and a ratio of $\approx~25~:~1$ between candidates and confirmed planets, which is equivalent to $0.2~\lesssim~N~\lesssim~1$. Our values are a little higher, probably due to the slightly larger FOV of $8^\circ~\times~8^\circ$ used in Fig. \ref{fig:N_50d} and due to the effect of blends and unresolved binaries (see discussion in Sect. \ref{sec:discussion}).

\begin{figure*}
  \centering
  \scalebox{0.51}{\includegraphics{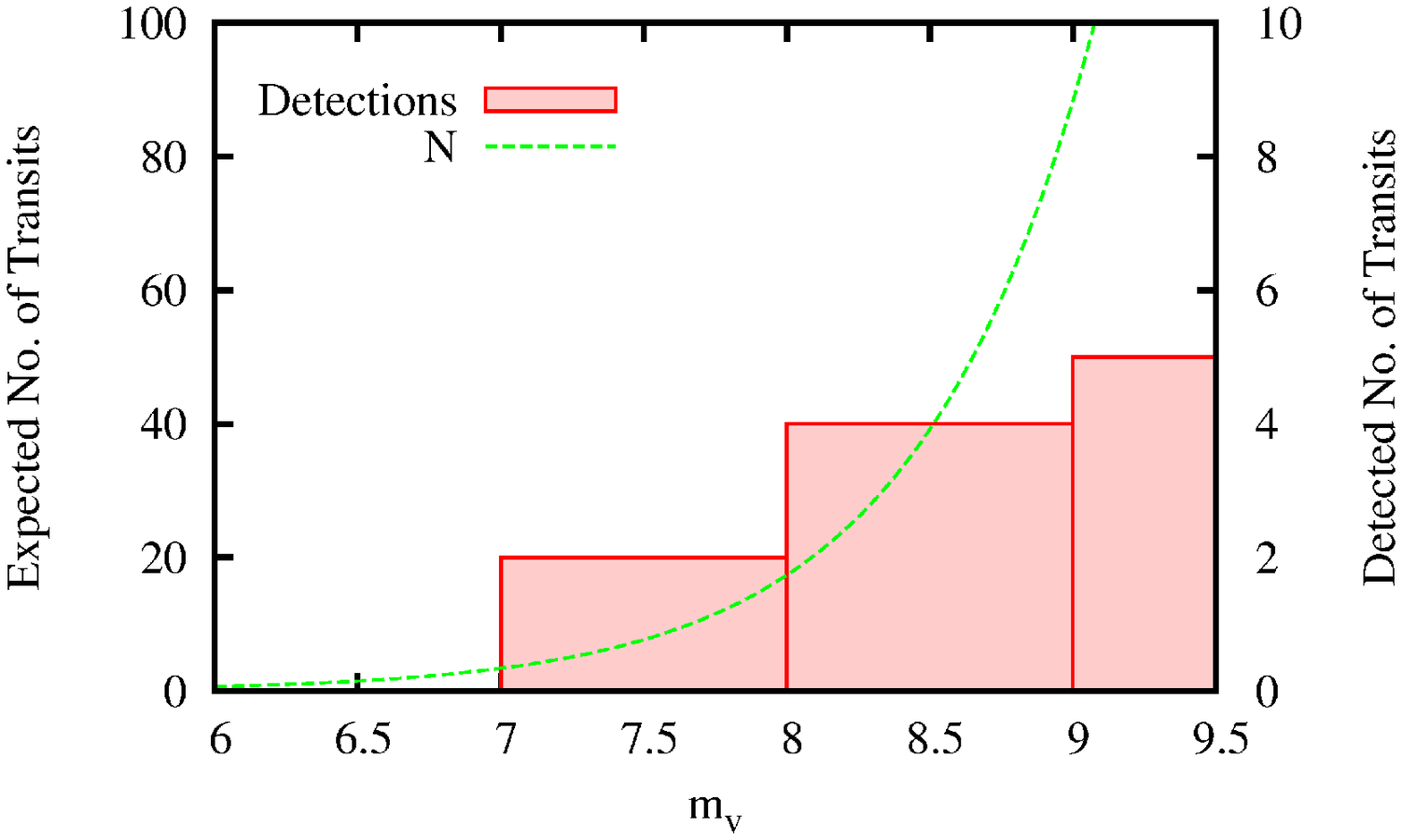}}
  \hspace{1.0cm}
  \scalebox{0.51}{\includegraphics{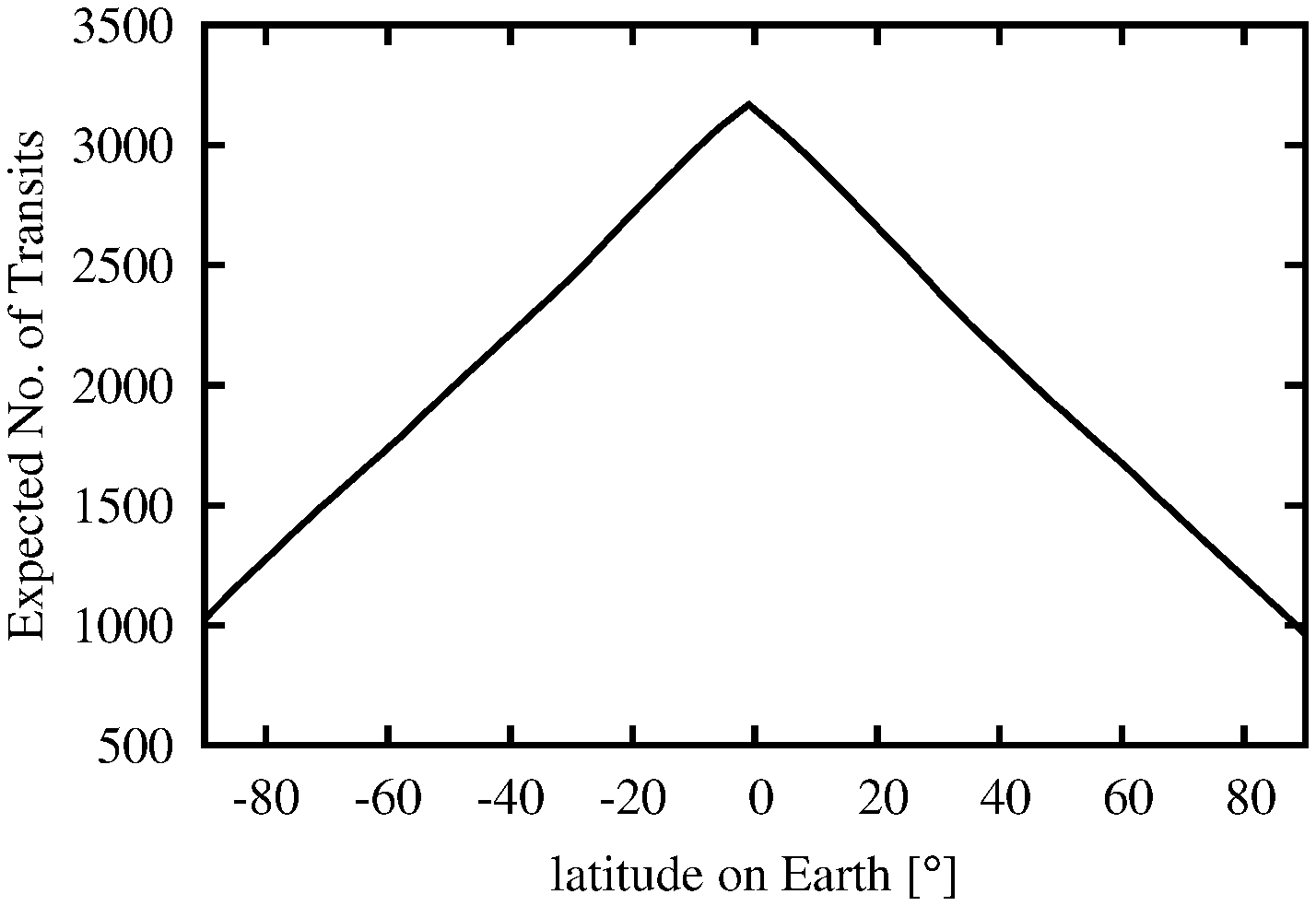}}
  \caption{Sky-integrated number of transits per magnitude (left panel) and as a function of latitude (right panel). \textit{Left}: While the green line represents our simulations, the rosy bars show the number of transiting planets per magnitude bin discovered so far. Note the different scales at the left and right ordinates! \textit{Right}: The triangle represents the expected number of transits that can be seen at elevations higher than $30^\circ$ over the horizon at a given latitude on Earth.}
  \label{fig:N_mag}
\end{figure*}

In the left panel of Fig. \ref{fig:N_mag}, we show the distribution of expected transits from our simulation as a function of the host stars' magnitudes compared to the distribution of the observed transiting exoplanets. The scales for both distributions differ about an order of magnitude, which is reasonable since only a fraction of actual transits is observed as yet. For $m_\mathrm{V}~<8^\mathrm{m}$, only HD209458b and HD189733b are currently know to show transits whereas we predict 20 of such transits with periods between 1.5\,d and 50\,d to occur in total. We also find that the number of detected transiting planets does not follow the shape of the simulated distribution for $m_\mathrm{V}~>~9^\mathrm{m}$. This is certainly induced by a lack of instruments with sufficient sensitivity towards higher apparent magnitudes, the much larger reservoir of fainter stars that has not yet been subject to continuous monitoring, and the higher demands on transit detection pipelines.

Our transit map allows us to constrain convenient locations for future ground-based surveys. A criterion for such a location is the number of transit events that can be observed from a given spot at latitude $l$ on Earth. To yield an estimate, we integrate $N$ over that part of the celestial plane that is accessable from a telescope situated at $l$. We restrict this observable fan to $l~-~60^\circ~<~\delta~<~l~+~60^\circ$, implying that stars with elevations $>~30^\circ$ above the horizon are observable. The number of the transit events with $m_\mathrm{V}~\lesssim~11.5^\mathrm{m}$ that is observable at a certain latitude on Earth is shown in the right panel of Fig. \ref{fig:N_mag}. This distribution resembles a triangle with its maximum almost exactly at the equator. Its smoothness is caused by the wide angle of $120^\circ$ that flattens all the fine structures that can be seen in Fig. \ref{fig:N_50d}.

\subsection{Transit detection}
\label{sub:detection}

So far, we have computed the sky and magnitude distributions of expected exoplanet transits with orbital periods between 1.5\,d and 50\,d, based on the stellar parameters from the Tycho data and empirical relations. In order to estimate if a possible transit can actually be observed, one also has to consider technical issues of a certain telescope as well as the efficiency and the selection effects of the data reduction pipelines. The treatment of the pipeline will not be subject of our further analysis. The relevant aspects for our concern are the pixel size of the CCD, its FOV, the $m_\mathrm{V}$ range of the CCD-telescope combination, and the declination fan that is covered by the telescope.

To detect a transit, one must be able to distinguish the periodic transit pattern within a light curve from the noise in the data. Since the depth of the transit curve is proportional to the ratio $A_P / A_\star$, where $A_P$ and $A_\star$ are the sky-projected areas of the planet and the star, respectively, and $R_P / R_\star = \sqrt{A_P / A_\star}$, with $R_P$ as the planetary radius, the detection probability for a certain instrument is also restricted to a certain regime of planetary radii. Assuming that the transit depth is about 1\,\%, the planetary radius would have to be larger than $\approx R_\star / 10$. We do not include an elaborate treatment of signal-to-noise in our considerations \citep[see][]{2007MNRAS.378..741A}. Since our focus is on MS stars and our assumptions for planetary occurrence are based on those of Hot Jupiters, our argumentation automatically leads to planetary transits of exoplanets close to $\approx R_\star / 10$.

We also do not consider observational aspects, such as integration time and an observer on a rotating Earth with observation windows and a finite amount of observing time \citep[see][for a review of these and other observational aspects]{2008MNRAS.386.1503F}. Instead, we focus on the technical characteristics of four well-established transit surveys and calculate the celestial distribution of expected exoplanet transit detections in principle by using one of these instruments. The impact of limited observing time is degraded to insignificance because the span of orbital periods we consider in Eq. (\ref{equ:Ni_expect}) reaches only up to $P_2 = 50$\,d. After repeated observations of the same field, such a transiting companion would be detected after $\lesssim~3$\,yr, which is the typical duty cycle of current surveys.

Our computations are compared for four surveys: BEST, XO, SuperWASP, and HATNet. This sample comprises the three most fruitful surveys in terms of first planet detections and BEST -- a search program that used a telescope instead of lenses. While observations with BEST have been ceased without any confirmed transit detection, XO has announced detections and SuperWASP and HATNet belong the most fruitful surveys to date. An overview of the relevant observational and technical properties of these surveys is given in Table \ref{tab:surveys}. For each survey, we first restrict the Tycho master sample to the respective magnitude range, yielding an $m_\mathrm{V}$-restricted sample. In the next step, we virtually observe the subsample with the fixed FOV of the survey telescope, successively grazing the whole sky with steps of $1^\circ$ between adjacent fields. The FOV is composed of a number of CCD pixels and each of these pixels contains a certain number of stars, whose combined photon fluxes merge into a count rate. Efficient transit finding has been proven to be possible from the ground in crowded fields, where target objects are not resolved from neighbor stars. To decide whether a hypothetical transit around the $i^{\mathrm{th}}$ star in the pixel would be detected, we simulate the effect of a transiting object that reduces the light flux contribution $l_i$ of the $i^{\mathrm{ht}}$ star on the combined flux $\sum_k^n l_k$ of the stars within a pixel. If the $i^\mathrm{th}$ magnitude variation on the pixel-combined light is ${\Delta}m_{\mathrm{V},i}~\geq~0.01^\mathrm{m}$, which is a typical accuracy limit of current ground-based surveys, then we keep this star for further analysis of the transit detection as described in Sects. \ref{sec:data_anal} and \ref{sub:occurance}, otherwise it is rejected. The fluxes, however, are not listed in the Tycho catalog; instead, we can use the visible magnitude $m_{\mathrm{V},i}$ of a star and calculate its relative flux $f_i/f_0$ with respect to a reference object with flux $f_0$ at magnitude $m_{\mathrm{V},0}$:

\begin{equation}
\frac{f_i}{f_0} = 10^{(m_{\mathrm{V},0} - m_{\mathrm{V},i})/2.5} \hspace{0.2cm} .
\end{equation}

\noindent
The magnitude variation can then be computed via

\begin{align}\label{equ:blening}
0.01 \stackrel{!}{\leq} \Delta m_{\mathrm{V},i} = & -2.5 \cdot \log\left( \frac{0.99 \cdot f_i + \displaystyle{\sum_{k \neq i}^n f_k}}{\displaystyle{\sum_k^n f_k}} \right)\\ \nonumber
= & -2.5 \cdot \log\left( \frac{0.99 \, \displaystyle{\cdot \, \frac{f_i}{f_0}} + \displaystyle{\sum_{k \neq i}^n \frac{f_k}{f_0}}}{\displaystyle{\sum_k^n \frac{f_k}{f_0}}} \right)  \hspace{0.2cm} .
\end{align}

\noindent
Without loss of generality we chose $m_{\mathrm{V},0} = 30^{\mathrm{m}}$ as reference magnitude.

\renewcommand{\arraystretch}{1.3}
\begin{table*}[h]
  \centering
  \caption{Instrumental properties of the treated surveys.}
  \label{tab:surveys}

    \begin{tabular}{c|c|r|c|c|c|c|c}

    \hline

    \hline

    \hline \hline

    {\sc Survey} & {\sc $\delta$ range} & {\sc FOV} & CCD Pixel Size & {\sc $m_\mathrm{V}$ Range} & {\sc $\delta$ \& $m_\mathrm{V}$} & {\sc $\delta$, $m_\mathrm{V}$ \& ${\Delta}m_\mathrm{V}$} & {\sc $\delta$, $m_\mathrm{V}$, ${\Delta}m_\mathrm{V}$ \& MS} \\
                 &                      &           &                &     {\sc [Mag.]}     & {\sc Limited Sample} &         {\sc Limited Sample}       &         {\sc Limited Sample}           \\
    \hline

    BEST      & $ -16^\circ< \delta < 90^\circ$ &  $3.1^\circ$ & \hspace{0.1cm} $5.5\,\arcsec/$mm & $8 < m_\mathrm{V} < 14$ & 546\,382 \ \tiny{(52.94\,\%)} & 516\,524 \ \tiny{(50.05\,\%)} & 222\,854 \ \tiny{(21.59\,\%)} \\

    XO        & $ -39^\circ< \delta < 90^\circ$ &  $7.2^\circ$ &      $25.4\,\arcsec/$mm           & $9 < m_\mathrm{V} < 12$ & 620\,477 \ \tiny{(60.12\,\%)} & 597\,842 \ \tiny{(57.93\,\%)} & 263\,213 \ \tiny{(25.51\,\%)} \\

    SuperWASP & $ -44^\circ< \delta < 90^\circ$ &  $7.8^\circ$ &      $13.8\,\arcsec/$mm          & $7 < m_\mathrm{V} < 12$ & 745\,227 \ \tiny{(72.21\,\%)} & 703\,707 \ \tiny{(68.19\,\%)} & 311\,404 \ \tiny{(30.18\,\%)} \\

    HATNet    & $ -28^\circ< \delta < 90^\circ$ &  $8.3^\circ$ &      $14.0\,\arcsec/$mm          & $7 < m_\mathrm{V} < 12$ & 721\,473 \ \tiny{(69.91\,\%)} & 686\,927 \ \tiny{(66.56\,\%)} & 283\,350 \ \tiny{(27.46\,\%)} \\

    \hline
  \end{tabular}

\flushleft
In the last three columns we list the reduced Tycho master sample of 1\,031\,992 stars after we applied the subsequent boundary conditions: the survey's sky-coverage ($\delta$ range), its $m_\mathrm{V}$ limitation, magnitude variation ${\Delta}m_\mathrm{V}~>~0.01^\mathrm{m}$ for the transit of a Jupiter-sized object around the $i^{\mathrm{th}}$ star in a pixel, and the boundary conditions for MS stars, for which the empirical relationships hold (see Sect. \ref{sub:derivation}). In braces we indicate the portion of the Tycho master sample.

\end{table*}

\section{Results}
\label{sec:results}

We develop a procedure for the calculation of the number of expected transit events to \textit{occur} around MS stars based on empirical relations between the stars and planets. This procedure is then applied to more than 1 million stars from the Tycho catalog to visualize the transit probability for all stars with $m_\mathrm{V}~\lesssim~11.5^\mathrm{m}$ as a sky map. We also compute the celestial distribution of the number of expected transit \textit{detections} for four different, well-established wide-field surveys.

\begin{figure*}
\begin{minipage}[t]{3cm}
  \scalebox{0.54}{\includegraphics{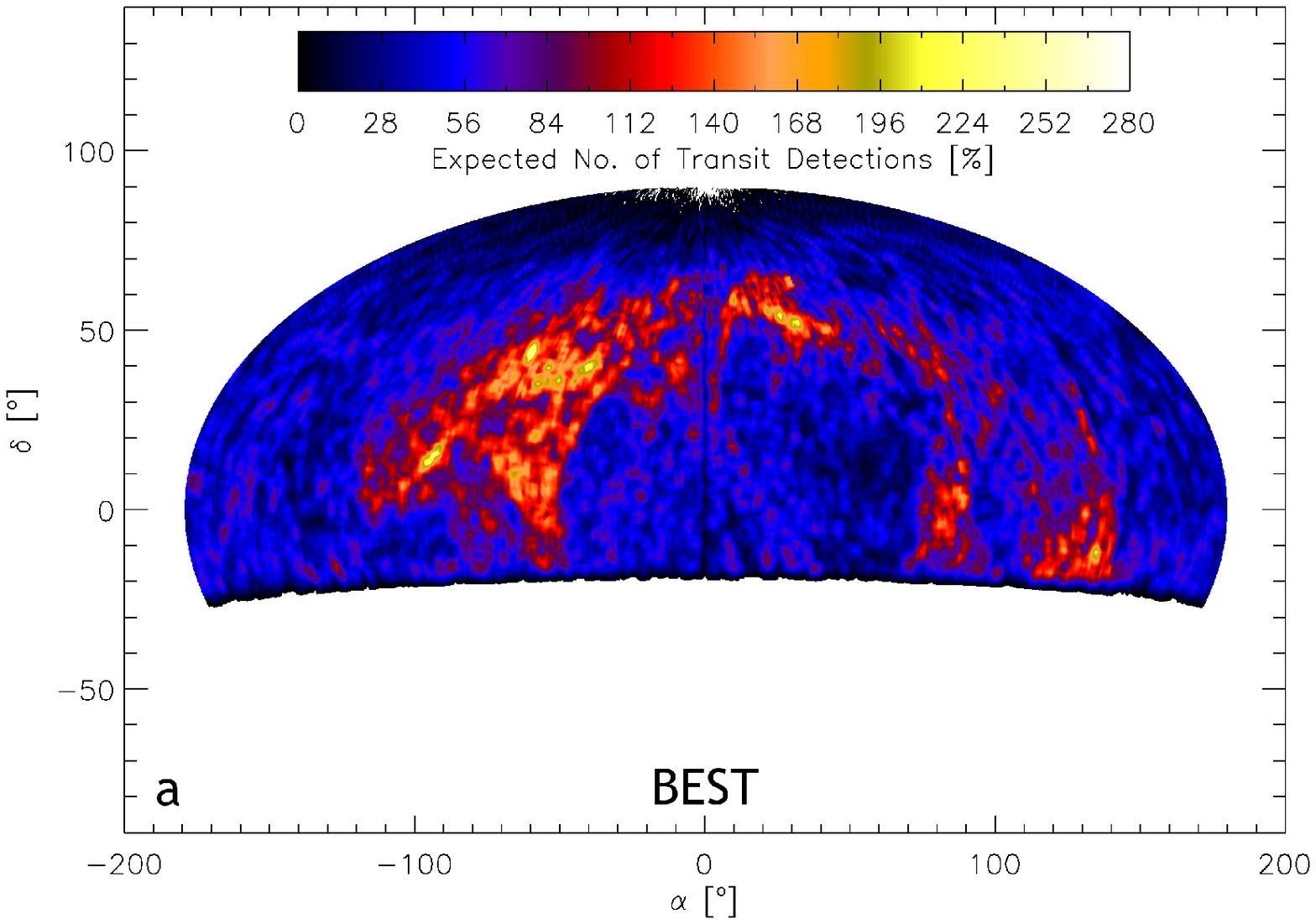}}
\end{minipage}
\hspace{6.3cm}
\begin{minipage}[t]{3cm}
  \scalebox{0.54}{\includegraphics{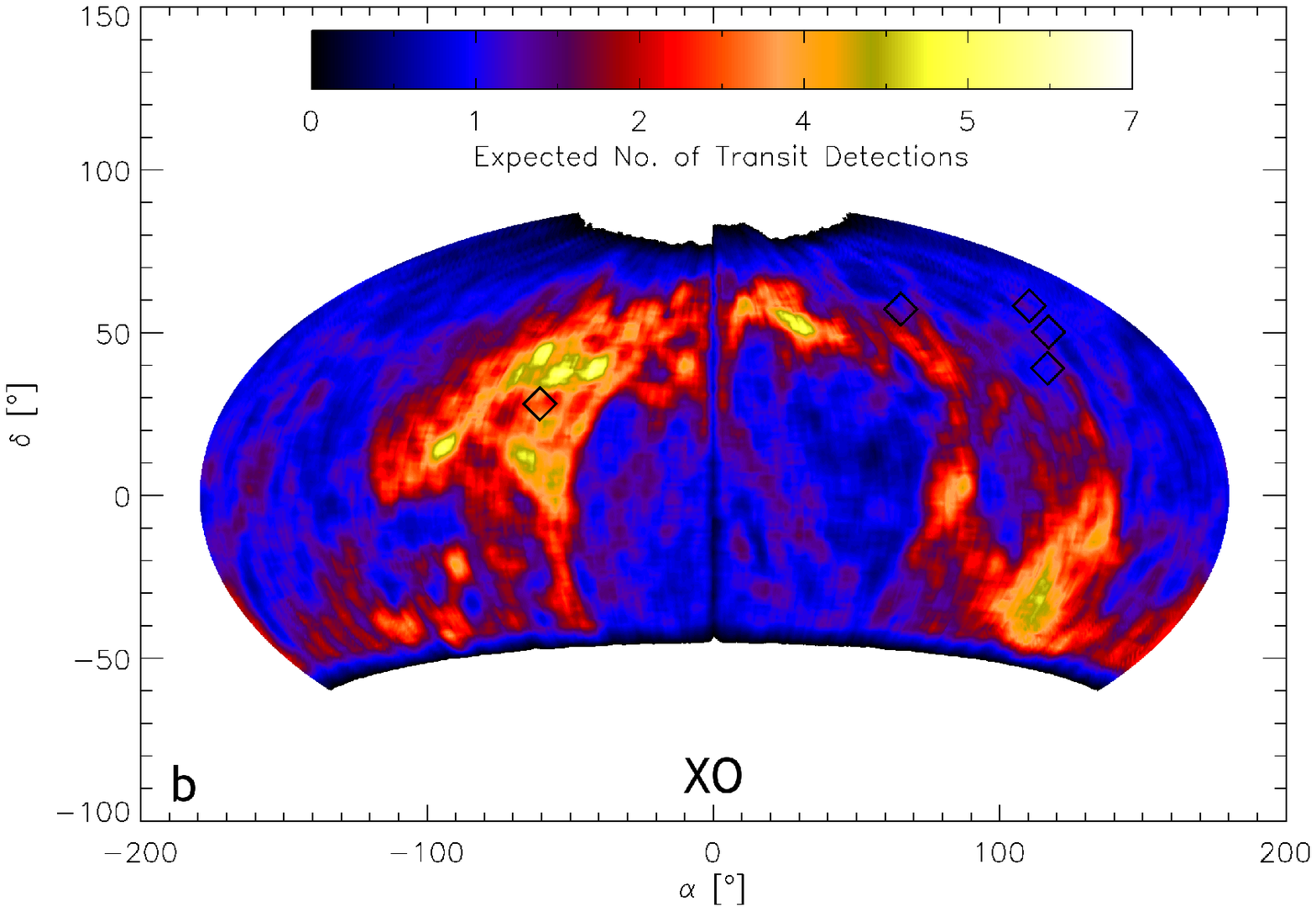}}
\end{minipage}\\
\begin{minipage}[t]{3cm}
  \vspace{0.3cm}
  \scalebox{0.54}{\includegraphics{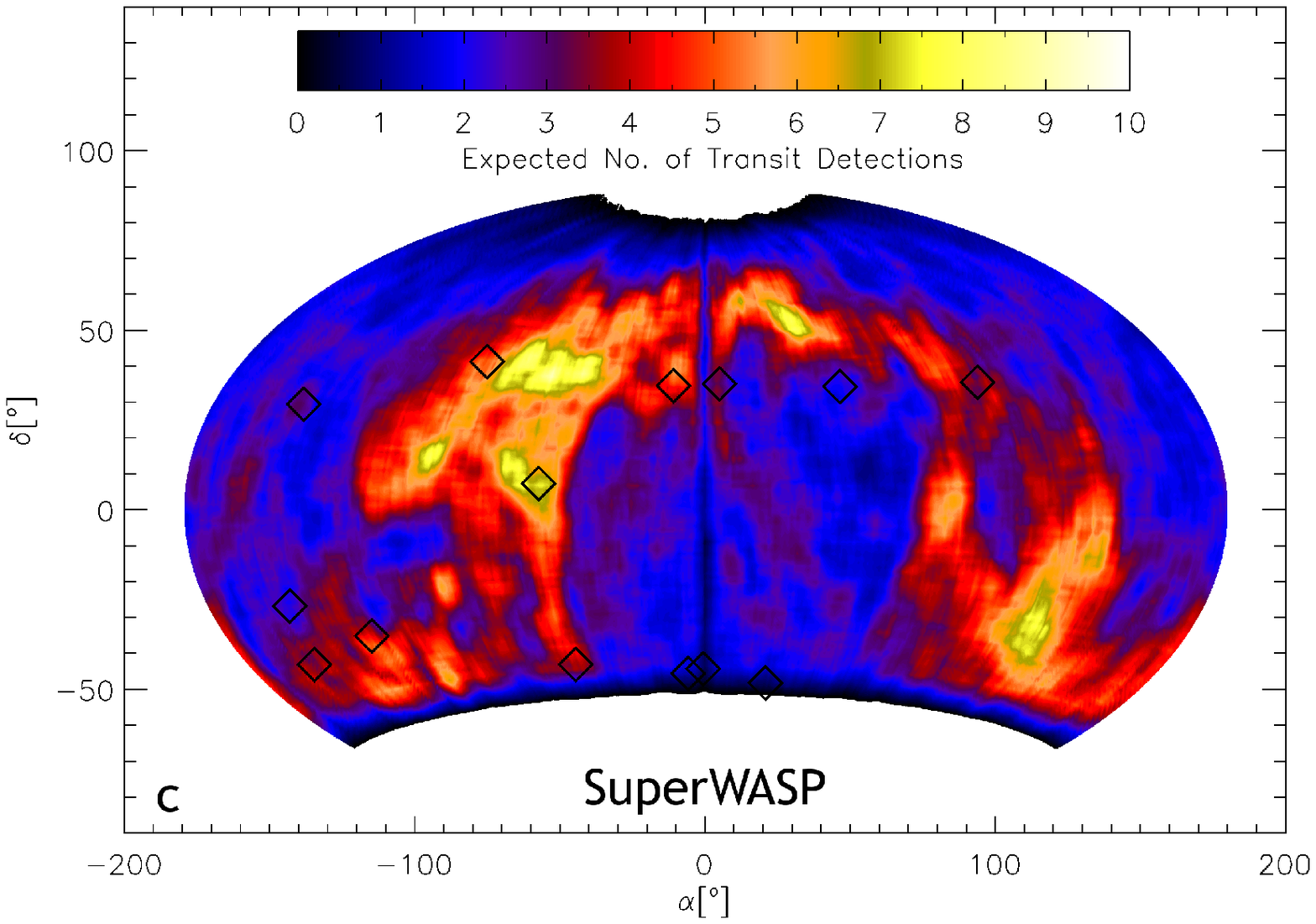}}
\end{minipage}
\hspace{6.3cm}
\begin{minipage}[t]{3cm}
  \vspace{0.3cm}
  \scalebox{0.54}{\includegraphics{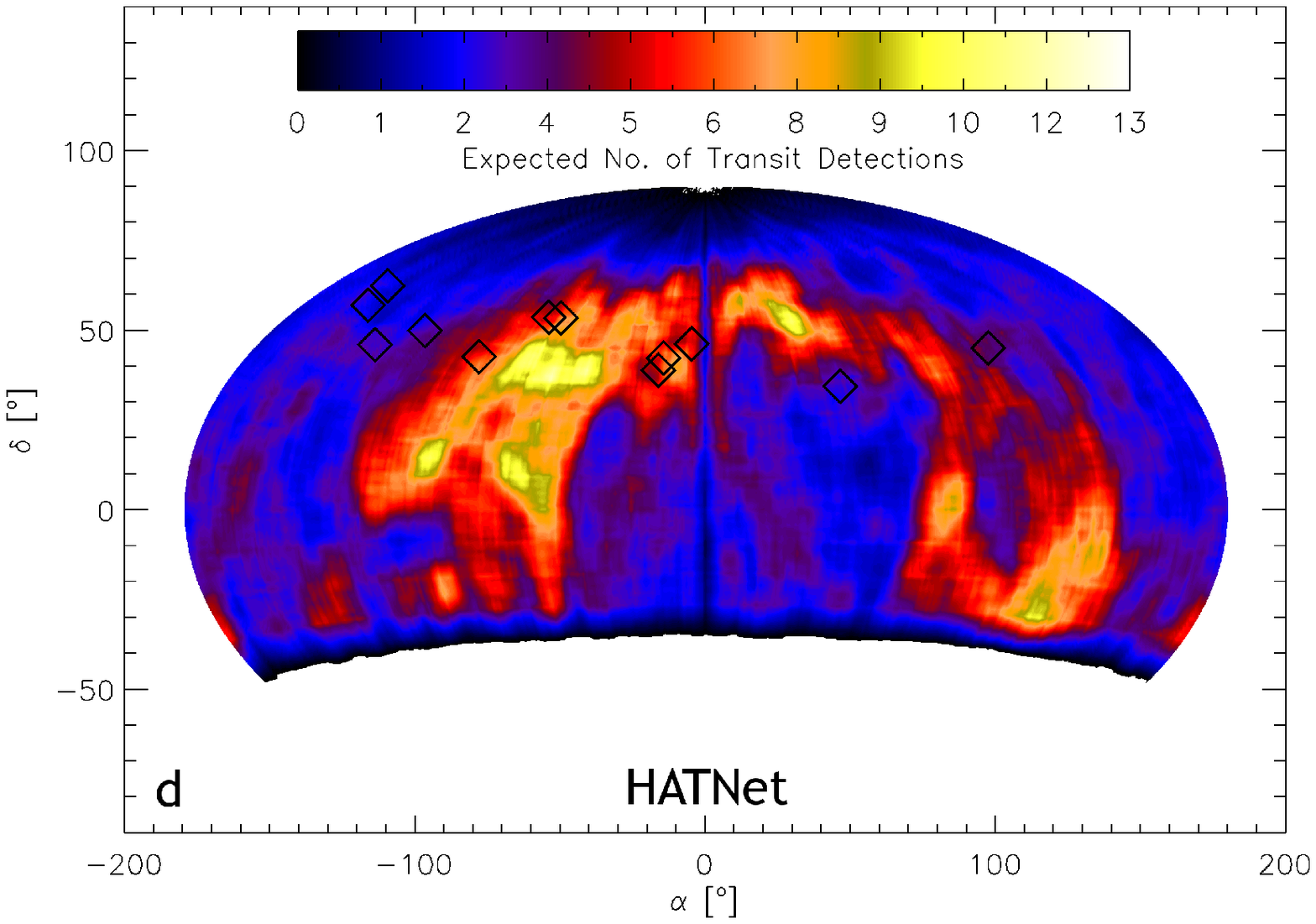}}
\end{minipage}
\caption{Sky maps with expected number of transit detections for BEST ({\bf a}), XO ({\bf b}), SuperWASP ({\bf c}), and HATNet ({\bf d}). Note the different scales of the color code! The published detections of the surveys, taken from the EPE as of September $1^\mathrm{st}$ 2009, are indicated with symbols.}
\label{fig:surveys}
\end{figure*}

In Fig. \ref{fig:surveys} we present the number of expected transit detections for the technical properties of BEST, XO, SuperWASP, and HATNet. As a general result from these maps, we find that the size of the FOV governs the detection efficiency of a camera. For the method applied here, the CCD resolution, i.e. the pixel size, has almost no impact since we neglect effects of noise, whereas in general the detection limits for transiting planets depend on the CCD resolution in terms of noise \citep{2005MNRAS.356..557K, 2005MNRAS.356.1466T, 2006MNRAS.373..231P}. In Table \ref{tab:surveys} you see that the restriction of ${\Delta}m_\mathrm{V}~>~0.01^\mathrm{m}$ almost doesn't reduce the sample. A large FOV, collecting the light of relatively many stars, outweighs a lower CCD resolution -- at least for the range of pixel sizes considered here. Even for the zones around and in the galactic center and anti-center where the stellar density increases drastically, the number of detectable transit events reaches its maximum. This was not foreseeable since blending, simulated by Eq. (\ref{equ:blening}), could have reduced the efficiency of transit detection within the crowded zones.

The four survey sky maps portray the very distinct efficiencies of the telescopes. The map of BEST reflects the stellar distribution of the Tycho data best due to the relatively high resolution of the CCD. However, the small FOV leads to very few expected transit detections. BEST's visible magnitude cut at the upper end is $14^\mathrm{m}$ while the Tycho catalog is complete only up to $11.5^\mathrm{m}$. Thus, a significant contribution of stars inside this range is excluded in Fig. \ref{fig:surveys}{\bf a}. BEST also covers the smallest portion of the sky, compared to the other surveys. The XO project yields a much more promising sky map, owed to the larger FOV of the lenses. But due to the relatively large pixel size and an adverse magnitude cut of $9~<~m_{\mathrm{V}}$, XO achieves lower densities of expected detections than SuperWASP and HATNet. As for SuperWASP and HATNet, the difference in the magnitude cuts with respect to the Tycho catalog is negligible for XO but tends to result in an underestimation of the expected detections. That part of the SuperWASP map that is also masked by HATNet looks very similar to the map of the latter one. While HATNet reaches slightly higher values for the expected number of detections at most locations, the covered area of SuperWASP is significantly larger, which enhances its efficiency on the southern hemisphere.

The total of expected transiting planets in the whole sky is 3412 (see Fig. \ref{fig:N_50d}). By summing up all these candidates within an observational fan of $30^\circ$ elevation above the horizon, we localize the most convenient site on Earth to mount a telescope for transit observations (see right panel in Fig. \ref{fig:N_mag}): it is situated at geographical latitude $l = -1^\circ$. Given that the rotation of the Earth allows a ground-based observer at the equator, where both hemispheres can be seen, to cover a larger celestial area than at the poles, where only one hemisphere is visible, this result could have been anticipated. Due to the non-symmetric distribution of stars, however, the shape of the sky-integrated number of expected transits as a function of latitude is not obvious. Figure \ref{fig:N_mag} shows that the function is almost symmetric with respect to the equator, with slightly more expected transits at the northern hemisphere. Furthermore, the number of expected transits to be observable at the equator is not twice its value at the poles, which is due to the inhomogeneous stellar distribution. In fact, an observer at the equator triples its number of expected transits with respect to a spot at the poles and can survey almost all of the 3412 transiting objects.

Based on the analysis of the magnitude distribution (left panel in Fig. \ref{fig:N_mag}), we predict 20 planets with $m_\mathrm{V} < 8$ to show transits with orbital periods between 1.5\,d and 50\,d, while two are currently known (HD209458b and HD189733b). These objects have proven to be very fruitful for follow-up studies such as transmission spectroscopy \citep{2002ApJ...568..377C, 2004ApJ...604L..69V, 2007Natur.447..183K, 2008ApJ...686..658S, 2008Natur.456..767G, 2008MNRAS.385..109P} and measurements of the Rossiter-McLaughlin effect \citep{Holt, 1924ApJ....60...15R, 1924ApJ....60...22M, 2006ApJ...653L..69W, 2007ApJ...665L.167W, 2007ApJ...667..549W, 2007PASJ...59..763N, 2008ApJ...683L..59C, 2009ApJ...700..302W}. Our analysis suggests that a significant number of bright transiting planets is waiting to be discovered. We localize the most promising spots for such detections.

\section{Discussion}
\label{sec:discussion}

Our values for the XO project are much higher than those provided by \citet{2008ApJ...686.1302B}, who also simulated the expected exoplanet transit detections of XO. This is due to their much more elaborate inclusion of observational constraints such as observational cadence, i.e. hours of observing per night, meteorologic conditions, exposure time, and their approach of making assumptions about stellar densities and the Galactic structure instead of using catalog-based data as we did. Given these differences between their approach and ours, the results are not one-to-one comparable. While the study of \citet{2008ApJ...686.1302B} definitely yields more realistic values for the expected number of transit detections considering all possible given conditions, we provide estimates for the celestial distribution of these detections, neglecting observational aspects.

In addition to the crucial respects that make up the efficiency of the projects, as presented in Table \ref{tab:surveys}, SuperWASP and HATNet benefit from the combination of two observation sites and several cameras, while XO also takes advantage of twin lenses but a single location. Each survey uses a single camera type and both types have similar properties, as far as our study is concerned. The transit detection maps in Fig. \ref{fig:surveys} refer to a single camera of the respective survey. The alliance of multiple cameras and the diverse observing strategies among the surveys \citep{2005PASP..117..783M, 2009IAUS..253...29C} bias the speed and efficiency of the mapping procedure. This contributes to the dominance of SuperWASP (18 detections, 14 of which have published positions)\footnote{EPE as of September $1^\mathrm{st}$ 2009} over HATNet (13 detections, all of which have published positions)\footnotemark[3], XO (5 detections, all of which have published positions)\footnotemark[3], and BEST (no detection)\footnotemark[3].

It is inevitable that a significant fraction of unresolved binary stars within the Tycho data blurs our results. The impact of unresolved binaries without physical interaction, which merely happen to be aligned along the line of sight, is significant only in the case of extreme crowding. As shown by \citet{2007ASPC..366..283G}, the fraction of planets not detected because of blends is typically lower than 10\,\%. The influence of unresolved physical binaries will be higher. Based on the empirical period distribution for binary stars from \citet{1991A&A...248..485D}, \citet{2008ApJ...686.1302B} estimate the fraction of transiting planets that would be detected despite the presence of binary systems to be $\approx~70$\,\%. Both the contribution of binary stars aligned by chance and physically interacting binaries result in an overestimation of our computations of $\approx~40$\,\%, which is of the same order as uncertainties arising from the empirical relationships we use. Moreover, as \citet{2006MNRAS.367.1103W} have shown, the density of eclipsing stellar binary systems increases dramatically towards the Galactic center. To control the fraction of false alarms, efficient data reduction pipelines, and in particular data analysis algorithms, are necessary \citep{2006MNRAS.365..165S}.

Recent evidence for the existence of ultra-short period planets around low-mass stars \citep{2009IAUS..253...45S}, with orbital periods $~<~1$\,d, suggests that we underestimated the number of expected transits to occur, as presented in Sect. \ref{sub:occurance}. The possible underestimation of exoplanets occurring at $\mathrm{[Fe/H]}_\star~<~0$ also contributes to a higher number of transits and detections than we computed here. Together with the fact that the Tycho catalog is only complete to $m_\mathrm{V} \lesssim 11.5^\mathrm{m}$, whereas the surveys considered here are sensitive to slightly fainter stars (see Table \ref{tab:surveys}), these trends towards higher numbers of expected transit detections might outweigh the opposite effect of unresolved binary stars.

A radical refinement of both our maps for transits occurrence and detections will be available within the next few years, once the `Panoramic Survey Telescope and Rapid Response System' (Pan-STARRS) \citep{2002SPIE.4836..154K} will run to its full extent. Imaging roughly 6000 square degrees every night with a sensitivity down to $m_\mathrm{V} \approx 24$, this survey will not only drastically increase the number of cataloged stars -- thus enhance our knowledge of the localization of putative exoplanetary transits -- but could potentially detect transits itself \citep{2009ApJ...704.1519D}. The Pan-STARRS catalog will provide the ideal sky map, on top of which an analysis presented in this paper can be repeated for any ground-based survey with the aim of localizing the most appropriate transit spots on the celestial plane. The bottleneck for the verification of transiting planets, however, is not the localization of the most promising spots but the selection of follow-up targets accessible with spectroscopic instruments. The advance to fainter and fainter objects thus won't necessarily lead to more transit confirmations. Upcoming spectrographs, such as the ESPRESSO{\MVAt}VLT and the CODEX{\MVAt}E-ELT \citep{2008PhST..130a4007P}, can be used to confirm transits around fainter objects. These next-generation spectrographs that will reveal Doppler fluctuations on the order of cm$\cdot$s$^{-1}$ will also enhance our knowledge about Hot Neptunes and Super-Earths, which the recently discovered transits of GJ\,436\,b \citep{2004ApJ...617..580B}, HAT-P-11\,b \citep{2009arXiv0901.0282B}, and CoRoT-7b \citep{2009arXiv0908.0241L} and results from \citet{2009IAUS..253..502L} predict to be numerous.

Further improvement of our strategy will emerge from the findings of more exoplanets around MS stars and from the usage of public data reservoirs like the NASA Star and Exoplanet Database\footnote{http://nsted.ipac.caltech.edu}, making assumptions about the metallicity distribution of planet host stars and the orbital period distribution of exoplanets more robust.

\begin{acknowledgements}

R.~Heller and D.~Mislis are supported by a PhD scholarship of the DFG Graduiertenkolleg 1351 ``Extrasolar Planets and their Host Stars''. We thank J. Schmitt, G. Wiedemann and M. Esposito  for their advice on the structure and readability of the paper. The referee deserves our honest gratitude for his comments on the manuscript which substantially improved the scientific quality of this study. This work has made use of Jean Schneider’s exoplanet database www.exoplanet.eu and of NASA's Astrophysics Data System Bibliographic Services.

\end{acknowledgements}

\bibliographystyle{aa} % style aa.bst
\bibliography{12378_proofs_corrected}

\end{document}